# Epitaxial Integration on Si(001) of Ferroelectric $Hf_{0.5}Zr_{0.5}O_2$ Capacitors with High Retention and Endurance

Jike Lyu, Ignasi Fina, Josep Fontcuberta, Florencio Sánchez*

Institut de Ciència de Materials de Barcelona (ICMAB-CSIC), Campus UAB, Bellaterra 08193, Barcelona, Spain

*Email: fsanchez@icmab.es

ABSTRACT Epitaxial ferroelectric $Hf_{0.5}Zr_{0.5}O_2$ films have been successfully integrated in a capacitor heterostructure on Si(001). The orthorhombic $Hf_{0.5}Zr_{0.5}O_2$ phase, [111] out-of-plane oriented, is stabilized in the films. The films present high remnant polarization $P_r$ close to 20 $\mu C/cm^2$, rivaling with equivalent epitaxial films on single crystalline oxide substrates. Retention time is longer than 10 years for writing field of around 5 MV/cm, and the capacitors show endurance up to $10^9$ cycles for writing voltage of around 4 MV/cm. It is found that the formation of the orthorhombic ferroelectric phase depends critically on the bottom electrode, being achieved on $La_{0.67}Sr_{0.33}MnO_3$ but not on $LaNiO_3$. The demonstration of excellent ferroelectric properties in epitaxial films of $Hf_{0.5}Zr_{0.5}O_2$ on Si(001) is relevant towards fabrication of devices that require homogeneity in the nanometer scale, as well as for better understanding of the intrinsic properties of this promising ferroelectric oxide.

1. INTRODUCTION

The discovery of robust ferroelectricity in the orthorhombic phase of HfO$_2$, a material fully compatible with CMOS processes, has been a breakthrough in the quest for nonvolatile memories.[1] The ferroelectric phase is usually stabilized in polycrystalline films where nanometric orthorhombic grains coexist with other dielectric phases.[2-4] The inherent inhomogeneity of polycrystalline films challenges the understanding of the intrinsic mechanisms of ferroelectricity in HfO$_2$ and its optimization. Moreover, oxygen vacancies form close to the interfaces with the commonly used TiN electrodes. It is normally accepted[5-7] that domain boundaries and oxygen vacancies have a prominent role on undesired properties as non-switchable ferroelectric domains, dielectric breakdown, electrical leakage, and wake-up effects. Thus, the polycrystalline nature of the HfO$_2$ films reported so far, challenges investigating the ultimate thickness limit of ferroelectricity in HfO$_2$ films. Remarkably, switchable polarization of around 3 μC/cm$^2$ has been measured in Hf$_{0.5}$Zr$_{0.5}$O$_2$ film as thin as 2.5 nm.[8] In spite of the progress, polycrystallinity can limit the performance and reliability of devices having small size of a few nanometers in thickness or a few tens of nanometers in a lateral dimension, and for these devices as well as for understanding the ferroelectric properties of HfO$_2$, epitaxial films are definitively required. But the progress in epitaxial ferroelectric HfO$_2$ is scarce in comparison with the huge progress achieved in polycrystalline films. Up to now, epitaxial stabilization of ferroelectric HfO$_2$ has been demonstrated on bare (001)-oriented yttria-stabilized zirconia (YSZ) substrates,[9-10] on YSZ(111) and YSZ(110) with indium tin oxide bottom electrodes,[11-13] and on SrTiO$_3$(001) with La$_{0.67}$Sr$_{0.33}$MnO$_3$ (LSMO) electrodes,[14-15] and very recently on YSZ-buffered Si(001) without electrodes.[16] The integration on Si(001) of epitaxial HfO$_2$ in a capacitor structure, i.e. with top and bottom electrodes as required for future integration in technology, has been elusive.

The epitaxial stabilization of a metastable phase depends critically on the substrate and buffer layers. For instance, the epitaxial stress can be responsible for the reported impact of the thickness on the amount of the orthorhombic phase in epitaxial HfO$_2$ films.[14-15] In the particular case of epitaxy on Si(001) substrates, thermal expansion mismatch can induce additional strain. We have investigated the stabilization of orthorhombic HfO$_2$ on epitaxial oxide electrodes deposited on Si(001), and we show here that ferroelectric capacitors with epitaxial Hf$_{0.5}$Zr$_{0.5}$O$_2$ (HZO) can be obtained on LSMO/LaNiO$_3$ (LNO)/CeO$_2$/YSZ/Si(001). In this heterostructure, YSZ is deposited first to provide an epitaxial template on Si(001). The CeO$_2$ and LNO layers are introduced to reduce progressively the lattice mismatch between LSMO and YSZ, making possible epitaxial growth of LSMO with [001] orientation. The orthorhombic HZO film grows epitaxially, with [111] out-of-plane texture, it has high remnant polarization close to 20 μC/cm$^2$, and presents retention time longer than 10 years for writing field of around 5 MV/cm and good endurance up to 10$^9$ cycles at around 4 MV/cm. These properties are similar to those obtained in HZO films of similar thickness on single crystalline oxide substrates.[14] Achievement of the epitaxial HZO films on Si(001) is a hallmark that can pave the way to better understand the properties of the orthorhombic phase in HfO$_2$-based compounds, as, for example, determining the minimum thickness preserving ferroelectricity[17-20] and using them in tunneling devices.

## 2. EXPERIMENTAL SECTION

HZO/LSMO/LNO/CeO$_2$/YSZ/Si(001) and HZO/LNO/CeO$_2$/YSZ/Si(001) heterostructures (sketched in Figure 1a) were fabricated in a single process by pulsed laser deposition (KrF excimer laser). The HZO films, of thickness t = 9.5 nm, were deposited at a laser repetition rate of 2 Hz. HZO, CeO$_2$ and YSZ were grown at substrate temperature (measured with a thermocouple inserted in the heater block) of 800 °C, and LSMO and LNO at 700 °C. The oxygen pressure was 0.1 mbar for deposition of HZO and LSMO, 0.15 mbar for LNO, and 4x10$^{-4}$ mbar for CeO$_2$ and YSZ. The heterostructures were cooled to room temperature under an oxygen pressure of 0.2 mbar. Crystal structure was characterized by X-ray diffraction (XRD) using Cu Kα radiation. θ-2θ scans were measured using a Siemens D5000 diffractometer equipped with point detector, and ϕ-scans using a Bruker D8-Advance diffractometer equipped with 2D detector. Atomic force microscopy (AFM) was used to study surface topography. Platinum circular top electrodes of diameter 20 μm were deposited using dc magnetron sputtering through stencil masks. Ferroelectric polarization loops, ferroelectric fatigue, and retention time were measured at room temperature in top-bottom configuration by means of an AixACCT TFAnalyser2000 platform. The leakage contribution to the polarization versus voltage loops has been minimized using the standard dynamic leakage current compensation (DLCC) procedure.[21-22]

## 3. RESULTS AND DISCUSSION

Figure 1b shows the XRD θ-2θ scans of the HZO/LSMO/ LNO/CeO$_2$/YSZ/Si(001) (upper scan) and HZO/LNO/CeO$_2$/YSZ/Si(001) (lower scan). The pattern of the HZO/LSMO/LNO/CeO$_2$/YSZ/Si(001) sample shows (00l) reflections corresponding to the Si wafer and the LSMO, LNO, CeO$_2$ and YSZ layers. There is also a peak at the position (2θ ~30°, see the zoom in the right panel) of the (111) reflection of the orthorhombic (o) phase of HZO. We have to note that the possible coexistence of minority tetragonal (t) phase cannot be discarded as t-HZO(101) and o-HZO(111) peaks would be overlapped. The tetragonal phase has been recently observed in epitaxial HZO films, being an ultrathin (thinner than 1 nm) interfacial layer between the HZO film and LSMO electrode.[14] The XRD 2θ-χ frame measured using a 2D detector (Figure S1b) shows a HZO(111) spot, which sharpness around χ = 0° points to epitaxial ordering. The 2θ-χ frame shows a low intensity m-HZO(002) spot, elongated along χ and located between the YSZ(002) and CeO$_2$(002) spots. Thus, minority monoclinic phase is present in HZO films that are predominantly orthorhombic and present (111) texture. The same crystal structure (phases coexistence and orientation) has been reported in HZO epitaxial films on LSMO/STO(001) [14-15] and (111) is also the preferential orientation observed in polycrystalline films on TiN electrodes.[2,23] It can be appreciated that the HZO(111) peak is slightly shifted towards lower angles in comparison with the corresponding position (marked with a vertical dashed line) of polycrystalline films.[5,24-25] The low intensity peaks close to HZO(111) are Laue reflections (a simulation is presented in Figure S2). Epitaxy of HZO and buffer layers was confirmed by XRD ϕ-scans (Figure 1c). The ϕ-scans around (111) reflections of Si, $CeO_2$, and YSZ show four peaks at the same ϕ angles, whereas the corresponding four peaks of LNO and LSMO are 45° apart of the former. The ϕ-scan around HZO(-111) presents four sets of three peaks, confirming the presence and crystal orientation of o-HZO, and indicating the presence of four crystal domains in o-HZO. The presence of these variants is due to the fact that HZO has grown (111) oriented (three-fold symmetry) on the LSMO(001) surface (four-fold symmetry). The epitaxial relationships are [1-10]HZO(111)/ [1-10]LSMO(001) /[1-10]LNO(001) /[100]$CeO_2$(001) /[100]YSZ(001) /[100]Si(001). The same epitaxial relationship has been reported very recently for HZO films on LSMO/$SrTiO_3$(001).[14-15] Wei et al.[14] found that HZO(111) films were compressively (tensile) strained in-the-plane (out-of-the-plane), which made the structure of the films rhombohedral. In the case of the epitaxial HZO(111) films on LSMO buffered Si(001), epitaxial strain either by thermal expansion differences or by lattice mismatch, would cause similar rhombohedral distortion. The topographic AFM image of this sample (Figure 1d) shows a very flat surface, with a root-means-square (rms) roughness 0.4 nm in the 5 μm x 5 μm scanned region. The surface flatness is also evidenced in the height profile (bottom panel) along the dashed line marked in the topographic image, with height variations of around only 1 nm along the 5 μm distance. We finally note that LSMO is critical in stabilizing the orthorhombic phase of HZO. Indeed, the XRD θ-2θ scan (Figure 1b) of the sample without this layer, HZO/LNO/$CeO_2$/YSZ/Si(001), does not show diffraction peaks of HZO. Only a low intensity elongated m-HZO(002) reflection is detected in XRD 2θ-χ frames measured using a 2D detector (Figure S1a), and very low intensity peaks corresponding to o-HZO(111), and (-111) and (002) of m-HZO were observed in grazing incidence XRD (Figure S3). This indicates a small amount of crystalline phases or very small crystal size when HZO is grown on LNO, in comparison to polycrystalline films obtained by annealing.[2] A set of HZO/LNO films was deposited on $SrTiO_3$(001) substrates to confirm the critical role of the electrode on the epitaxial stabilization of o-HZO. In agreement with the samples on Si(001), equivalent HZO films on LNO/$SrTiO_3$(001), not shown here, did not contain the orthorhombic phase.

Ferroelectric characterization of HZO/LSMO/LNO/$CeO_2$/YSZ/Si(001) and HZO/LNO/ $CeO_2$/YSZ/Si(001) samples is shown in Figure 2a,b, respectively. The polarization versus voltage loop recorded for the HZO sample growth on LSMO (Figure 2a) shows a clear ferroelectric hysteresis loop. The remnant polarization is 18 μC/$cm^2$, and the average coercive voltage 3.2 V (3.3 MV/cm) with an imprint voltage displacing the loop by around 0.6 V (620 kV/cm) towards negative voltage. It signals the presence of an internal field pointing from top Pt towards bottom LSMO electrode. The loop is saturated as demonstrated by the increasing voltage loops shown in Figure S4. It is remarkable that the ferroelectric switching is observed in pristine devices with no need of wake-up electric cycling.[26] On the other hand, the presence of a large dielectric contribution is manifested by the substantial slope of the loop.[14-15,24-25] In Figure 2a we show (red curve) the compensated loop after removing the electric susceptibility contribution by linear substraction of the constant slope (corresponding to $\varepsilon_r = 33$) in the pristine loop. Note that both loops show a small contribution of leakage current and series resistance at high applied voltage as denoted by the small loop aperture of the loop at high voltage. This results in an overestimation of polarization of about 1 µC/cm$^2$ (Figure S5). The remnant polarization of 18 µC/cm$^2$ is comparable to the highest values for polycrystalline HZO films,[2,26] and is similar to that of epitaxial films on perovskite SrTiO$_3$(001) substrates of same thickness.[14-15] The polarization is well below the saturation polarization of 53 µC/cm$^2$ theoretically predicted for this compound.[27] A reason for the smaller value is that the polar axis of orthorhombic hafnia is [010], whereas the epitaxial films[14-15] are [111] oriented along the out-of-plane direction. Moreover, the lattice strain in epitaxial films (in particular they show expanded (111) interplanar spacing compared to polycrystalline films) can influence the polarization. Also, paraelectric monoclinic crystallites could be present in the film as secondary phase, as found in similar HZO films on LSMO/ SrTiO$_3$(001).[14] Finally, some domains could be non-switchable due to pinning by defects. On the other hand, we note that in sharp contrast with the film on the LSMO electrode, when HZO is grown directly on the LNO electrode (Figure 2b), the polarization versus voltage measurement does not show any signature of ferroelectricity. The small aperture is attributed to leakage current.[28-29] Measurement in top-top configuration confirmed absence of signatures of ferroelectricity (Figure S6).

Retention experiments are shown in Figure 3 for several applied poling voltages. It can be observed that if polarization is settled by voltage pulses of 5.0 and 5.5 V, extrapolated data safely overpasses 10 years retention. A high polarization window of 2P$_r$ around 14 µC/cm$^2$ is extrapolated 10 years after 5.5 V poling. The experiments performed at 4.5 and 4.0 V show that retention reaches 10 and 1 year, respectively. Measurements shown in Figure S7 reveal that none of the two opposite ferroelectric states are favored by the presence of the imprint field observed in Figure 2.

Endurance characterization is summarized in Figure 4. Polarization loops collected either in the pristine state and after cycling the sample for $10^3$ and $10^5$ times using an applied voltage of 5.0 V are shown in Figure 4a. When the cycling number exceeds $10^5$ cycles at 5.0 V, dielectric rupture occurs. The dielectric rupture is delayed up to $10^8$ cycles when reducing the switching voltage (4.5 V) is used (Figure 4b)). If switching voltage is further decreased (4.0 V) ruptures does not occur in the explored cycling range (Figure 4c); however remnant polarization is smaller and the polarization window is reduced to 2P$_r$ = 2 µC/cm$^2$ after $10^9$ cycles. Figure 4d summarizes the effect of electric cycling on the remnant polarization normalized to its initial value (P$_0$). These data show that the polarization fatigue effect is independent of the maximum applied voltage and thus only caused by ferroelectric switching. Therefore, the maximum applied voltage determines the dielectric rupture but not the polarization reduction.

Leakage current curves are shown in Figure 5. It has to be noted that some polycrystalline ferroelectric HfO$_2$ films have been reported having smaller leakage[30] in spite of the presence of grain boundaries. Epitaxial films are not necessarily monocrystalline and they can present grain boundaries too. Indeed, the presence of four o-HZO(111) crystal variants imply domain walls.

Further studies are needed to determine the origin of the leakage in epitaxial films. The leakage can be an important factor causing fatigue, so to further investigate the endurance, leakage current density measurements were performed after different cycling number as shown in Figures 5a, 5b, and 5c for maximum applied voltage of 5.0, 4.5 and 4.0 V. It can be observed that leakage current remains almost constant up to $10^4$, $10^6$, $10^6$ cycles for maximum applied voltage of 5.0, 4.5 and 4.0 V, respectively. Afterwards, the leakage current suddenly increases (marked by an arrow). This effect can be clearly inferred in Figure 5d, where the leakage current at 300 kV/cm is plotted as a function of cycling number. It is observed that the leakage increases significantly after around $10^4$ cycles at 5.0 V and after $10^7$ cycles at 4.5 or 4.0 V. In spite of the increased leakage, the polarization window was above 2 μC/cm$^2$ up to $10^8$ and $10^9$ cycles at 4.5 and 4.0 V, respectively. Thus, the epitaxial HZO films on Si(001) present good endurance in addition of long retention time. The films, epitaxially grown on oxide LSMO electrodes, offer opportunities to develop devices like ferroelectric tunnel junctions requiring high homogeneity and accurate control of the interfaces.

4. CONCLUSIONS

In conclusion, ferroelectric capacitors with epitaxial Hf$_{0.5}$Zr$_{0.5}$O$_2$ have been integrated on Si(001) by the first time. The epitaxial integration is achieved on a buffer heterostructure that includes YSZ as epitaxial template. The epitaxial films on Si(001) reach the ferroelectric properties achieved in epitaxial films of similar thickness on oxide single crystalline substrates, with remnant polarization close to 20 μC/cm$^2$. Cycling electric field dependences of retention and endurance have been determined, showing retention time longer than 10 years for writing field of around 5 MV/cm and endurance up to $10^9$ cycles at around 4 MV/cm. Epitaxial films, without the inherent inhomogeneity present in polycrystalline films, can make easier prototyping tunnel devices and investigating intrinsic ferroelectric properties.

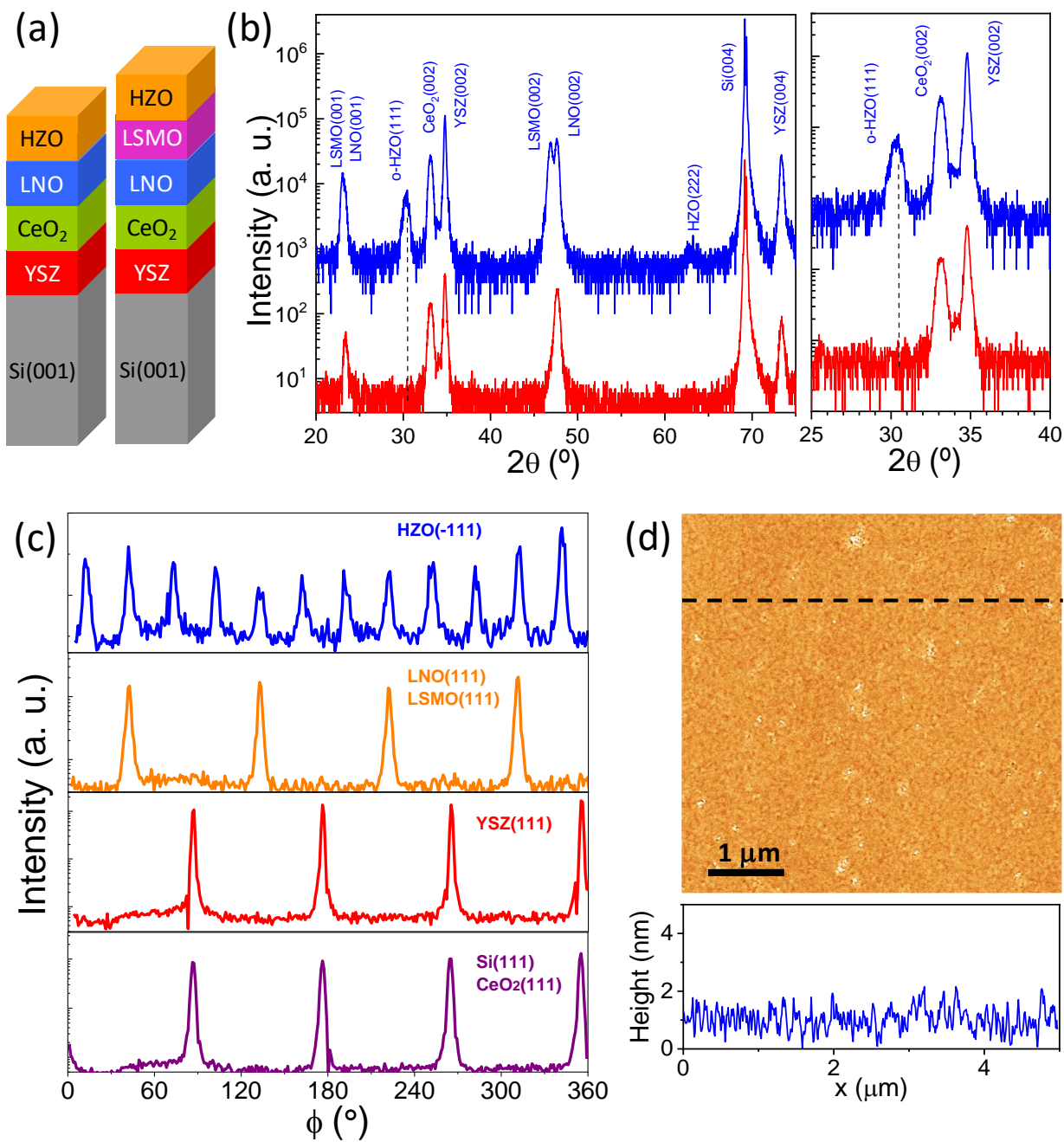

**Figure 1.** (a) Sketch of the epitaxial heterostructures. (b) XRD θ-2θ scans of the HZO/LSMO/LNO/CeO$_2$/YSZ/Si(001) (upper scan) and HZO/LNO/CeO$_2$/YSZ/Si(001) (lower scan). A zoomed region around the HZO(111) reflection is shown in the right panel. (c) ϕ-scans around the o-HZO(-111) and (111) reflections of the LSMO, LNO, CeO2, YSZ and Si corresponding to the HZO/LSMO/LNO/CeO$_2$/YSZ/Si(001) sample. (d) Topographic AFM image of the HZO/LSMO/LNO/CeO$_2$/YSZ/Si(001) sample. Bottom panel: height profile along the horizontal dashed line.

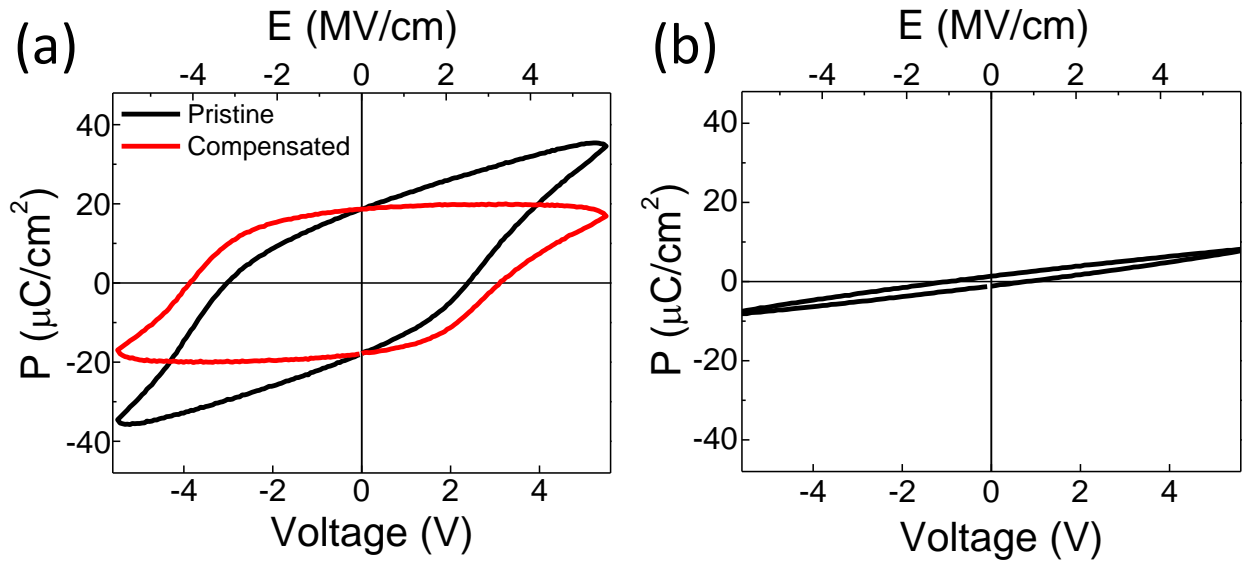

**Figure 2**. (a) Ferroelectric hysteresis loop (black curve) at 1 kHz for HZO/LSMO/LNO/CeO$_2$/YSZ/Si(001) sample. Compensated loop (removing the electric susceptibility contribution of HZO) by substraction of the linear slope is also included (red curve). (b) Ferroelectric hysteresis loop, obtained at 1 kHz without using the DLCC method, for HZO/LNO/CeO$_2$/YSZ/Si(001) sample.

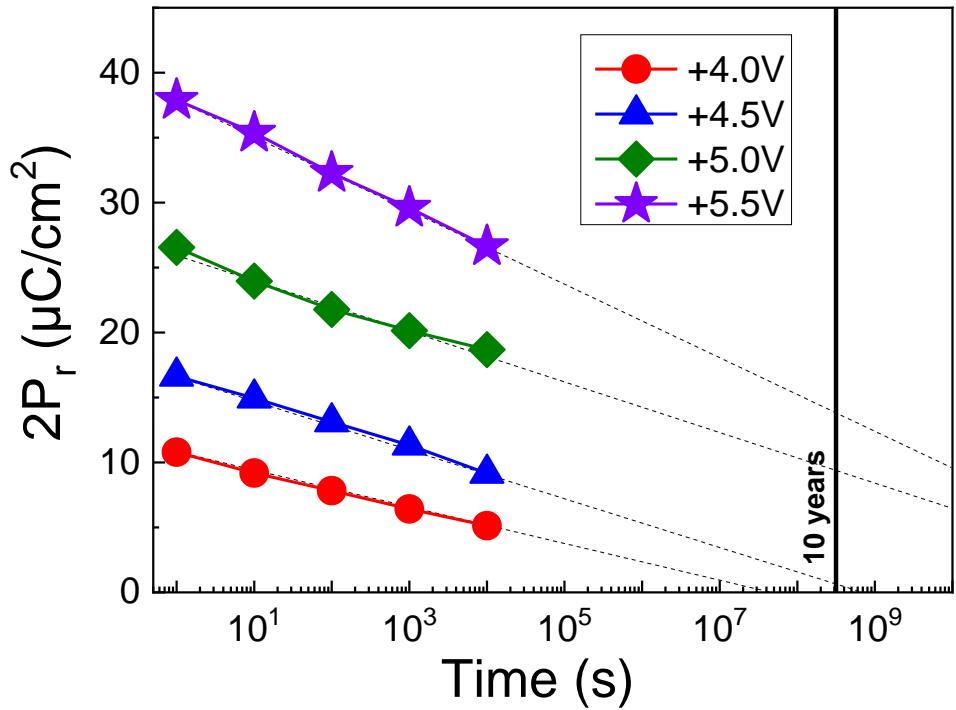

**Figure 3.** Ferroelectric retention (2P$_r$) after poling the film with voltage pulses of the indicated amplitude. Reading voltage was the same that used for writing. Dashed lines are guides to the eye. Vertical solid line marks time of 10 years.

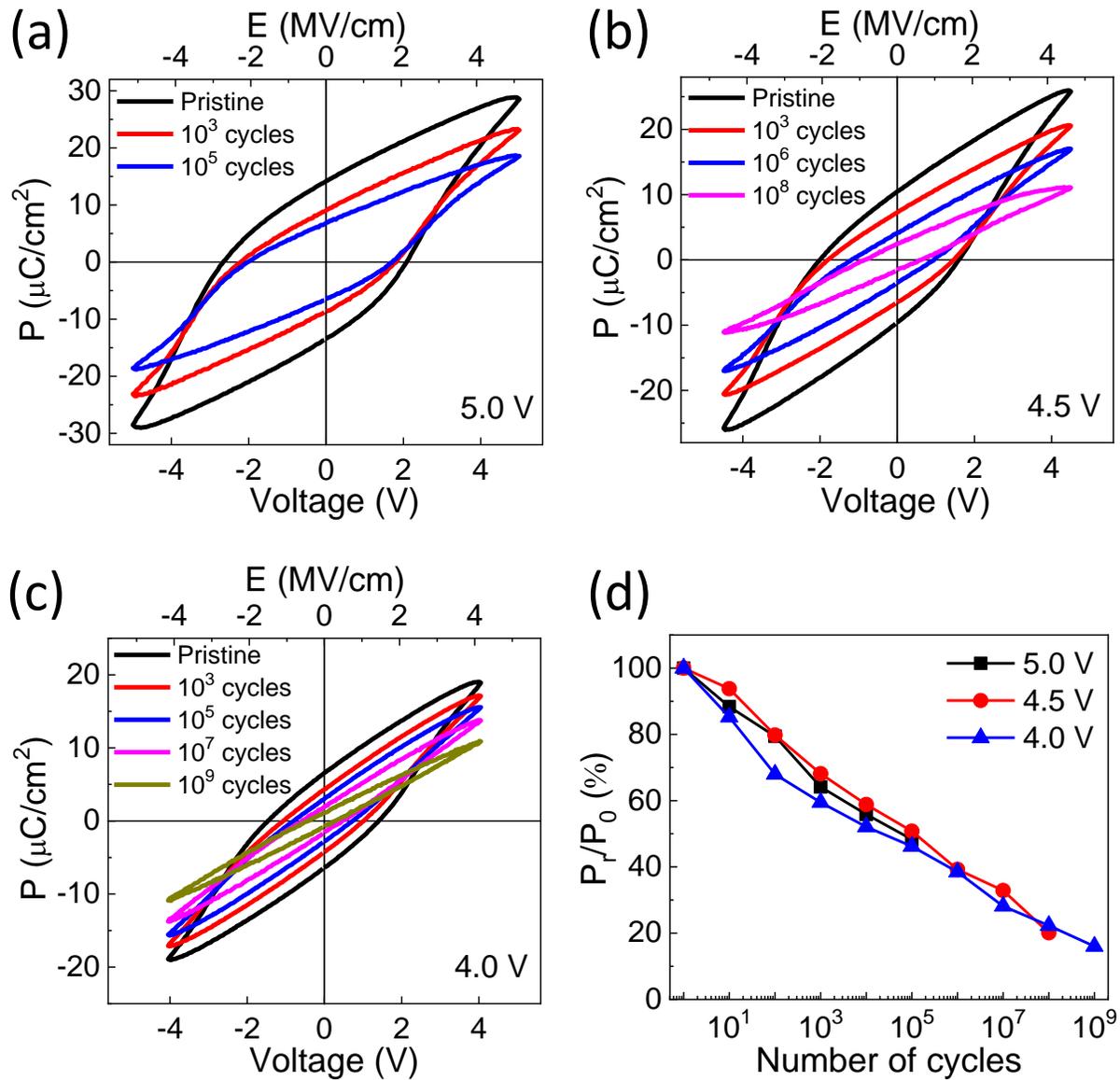

**Figure 4.** (a, b, c) Polarization versus voltage loops measured for the pristine state and after indicated number of electric cycles of 5.0, 4.5 and 4.0 V at 10 kHz, respectively. (d) Summary of the remnant polarization (positive and negative average values) normalized to its initial value ($P_0$) obtained in (a, b, c) versus cycling number for the different maximum applied voltage.

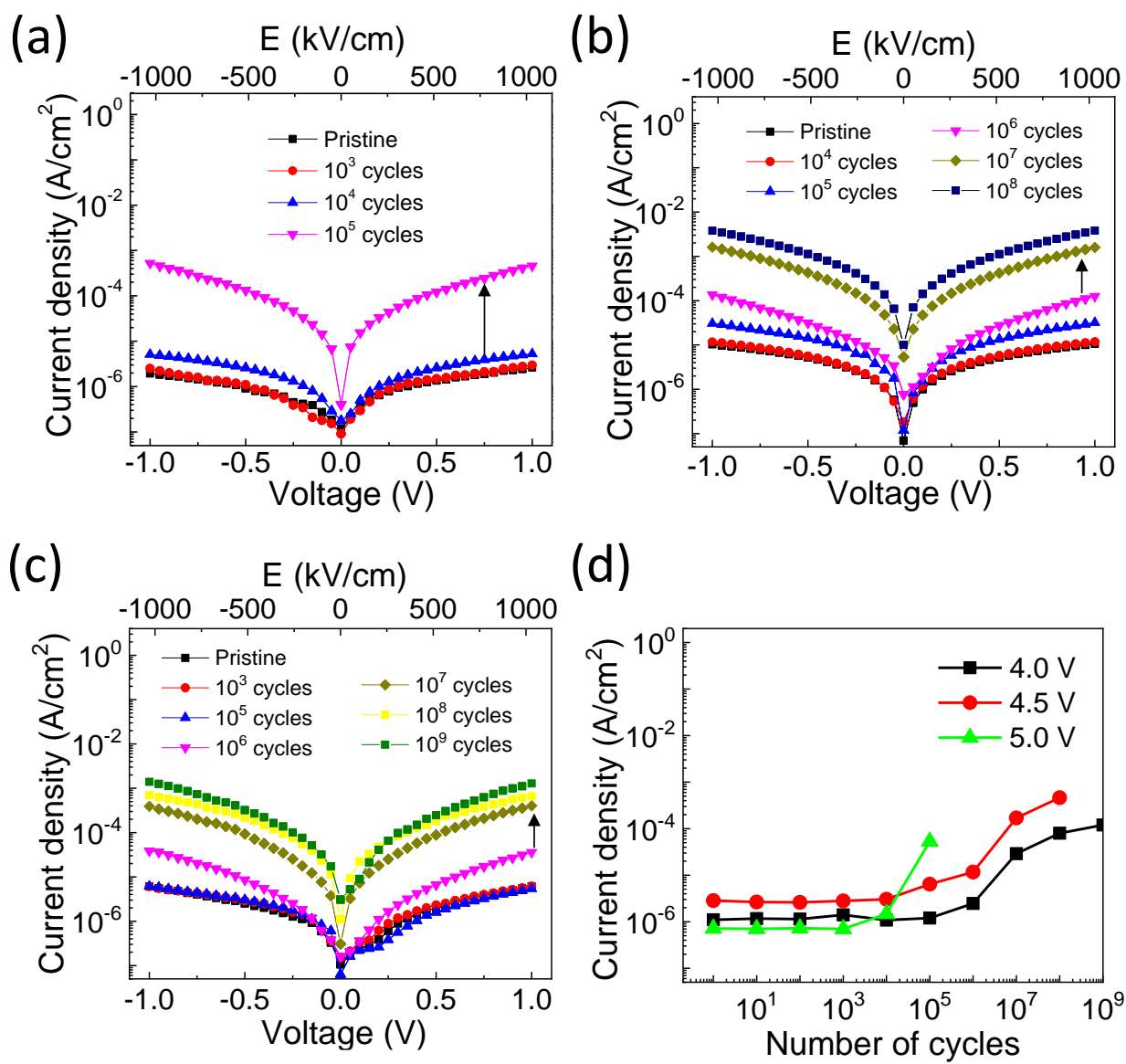

**Figure 5.** (a, b, c) Current leakage density versus electric field characteristics of the pristine state and after the indicated number of bipolar cycles at 5.0, 4.5 and 4.0 V, respectively. (d) Current leakage density evaluated at 300 kV/cm versus the number of bipolar cycles.

ASSOCIATED CONTENT

**Supporting Information**. XRD 2θ-χ frames, recorded with a 2D detector. XRD θ-2θ scan of the HZO/LSMO/LNO/ CeO$_2$/YSZ/Si(001) sample with simulations of Laue reflections around HZO(111) and CeO$_2$(002) reflections. Grazing incidence XRD scan of the HZO/LNO/CeO$_2$/

YSZ/Si(001) sample. Electrical measurements at high voltage in the HZO/LNO/CeO$_2$/YSZ/Si(001) sample confirming absence of ferroelectricity. Current versus voltage hysteresis loops recorded at increasing voltage in the HZO/LSMO/LNO/CeO$_2$/YSZ/Si(001) sample. Ferroelectric retention in the HZO/LSMO/LNO/CeO$_2$/YSZ/Si(001) sample after positive and negative poling voltage pulses. Compensation of leakage in ferroelectric hysteresis loops.

**Corresponding Author**

*Email: fsanchez@icmab.es

ACKNOWLEDGMENT

Financial support from the Spanish Ministry of Science, Innovation and Universities, through the "Severo Ochoa" Programme for Centres of Excellence in R&D (SEV-2015-0496) and the MAT2017-85232-R (AEI/FEDER, EU), MAT2014-56063-C2-1-R, and MAT2015-73839-JIN projects, and from Generalitat de Catalunya (2017 SGR 1377) is acknowledged. I.F. acknowledges RyC contract RYC-2017-22531. JL is financially supported by China Scholarship Council (CSC) with No. 201506080019. JL work has been done as a part of his Ph.D. program in Materials Science at Universitat Autònoma de Barcelona.

REFERENCES

(1) Boscke, T. S.; Müller, J.; Bräuhaus, D.; Schröder, U.; Böttger, U. Ferroelectricity in Hafnium Oxide Thin Films. *Appl. Phys. Lett.* **2011**, 99, 102903.

(2) Park, M. H.; Lee, Y. H.; Kim, H. J.; Kim, Y. J.; Moon, T.; Kim, K. D.; Müller, J.; Kersch, A.; Schroeder, U.; Mikolajick, T.; Hwang, C. S.. Ferroelectricity and Antiferroelectricity of Doped Thin HfO$_2$-Based Films. *Adv. Mater.* **2015**, 27, 1811-1831.

(3) Fan, Z.; Chen, J.; Wang, J. Ferroelectric HfO$_2$-based Materials for Next-Generation Ferroelectric Memories. *J. Adv. Dielectrics* **2016**, 6, 1630003.

(4) Park, M. H.; Lee, Y. H.; Kim, H. J.; Kim, Y. J.; Moon, T.; Kim, K. D.; Hyun, S. D.; Mikolajick, T.; Schroeder, U.; Hwang, C. S. Understanding the Formation of the Metastable Ferroelectric Phase in Hafnia–Zirconia Solid Solution Thin Films. *Nanoscale* **2018**, 10, 716-725.

(5) Fengler, F.P.G.; Nigon, R.; Muralt, P.; Grimley, E.D.; Sang, X.; Sessi, V.; Hentschel, R.; LeBeau, J.M.; Mikolajick, T.; Schroeder, U. Analysis of Performance Instabilities of Hafnia-Based Ferroelectrics Using Modulus Spectroscopy and Thermally Stimulated Depolarization Currents. *Adv. Electron. Mater.* **2018**, 4, 1700547.

(6) Fengler, F. P. G.; Pešić, M.; Starschich, S.; Schneller, T.; Künneth, C.; Böttger, U.; Mulaosmanovic, H.; Schenk, T.; Park, M. H.; Nigon, R.; Muralt, P.; Mikolajick, T.; Schroeder, T. Domain Pinning: Comparison of Hafnia and PZT Based Ferroelectrics. *Adv. Electr. Mater.* **2017**, 3, 1600505.

(7) Pešić, M.; Hoffmann, M.; Richter, C.; Mikolajick, T.; Schroeder, U. Nonvolatile Random Access Memory and Energy Storage Based on Antiferroelectric Like Hysteresis in $ZrO_2$. *Adv. Funct. Mater.* **2016**, 26, 7486-7494.

(8) Chernikova, A.; Kozodaev, M.; Markeev, A.; Negrov, D.; Spiridonov, M.; Zarubin, S., Bak, O.; Buragohain, P.; Lu, H.; Suvorova, E.; Gruverman, A.; Zenkevich, A. Ultrathin $Hf_{1/2}Zr_{1/2}O_2$ Ferroelectric Films on Si. *ACS Appl. Mater. Interf.* **2016**, 8, 7232-7237.

(9) Shimizu, T.; Katayama, K.; Kiguchi, T.; Akama, A.; Konno, T. J.; Funakubo, H. Growth of Epitaxial Orthorhombic $YO_{1.5}$-substituted $HfO_2$ Thin Film. *Appl. Phys. Lett.* **2015**, 107, 032910.

(10) Katayama, K.; Shimizu, T.; Sakata, O.; Shiraishi, T.; Nakamura, S.; Kiguchi, T.; Akama, A.; Konno, T.J.; Uchida, H.; Funakubo, H. Orientation Control and Domain Structure Analysis of {100}-Oriented Epitaxial Ferroelectric Orthorhombic $HfO_2$-based Thin Films. *J. Appl. Physics* **2016**, 119, 134101.

(11) Mimura, T.; Katayama, K.; Shimizu, T.; Uchida, H.; Kiguchi, T.; T. Akama, T.; Konno, T. J.; Sakata, O.; Funakubo, H. Formation of (111) Orientation-Controlled Ferroelectric Orthorhombic $HfO_2$ Thin Films from Solid Phase via Annealing. *Appl. Phys. Lett.* **2016**, 109, 052903.

(12) Katayama, K.; Shimizu, K.; Sakata, O.; Shiraishi, T.; Nakamura, S.; Kiguchi, T.; Akama, A.; Konno, T.J.; Uchida, H.; Funakubo, H. Growth of (111)-Oriented Epitaxial and Textured Ferroelectric Y-doped $HfO_2$ Films for Downscaled Devices. *Appl. Phys. Lett.* **2016**, 109, 112901.

(13) Shimizu, T.; Katayama, K.; Kiguchi, T.; Akama, A.; Konno, T. J.; Sakata, O.; Funakubo, H. The Demonstration of Significant Ferroelectricity in Epitaxial Y-doped $HfO_2$ Film. *Sci. Rep.* **2016**, 6, 32931.

(14) Wei, Y.; Nukala, P.; Salverda, M.; Matzen, S.; Zhao, H. J.; Momand, J.; Everhardt, A.; Blake, G. R.; Lecoeur, P.; Kooi, B. J.; Íñiguez, J.; Dkhil, B.; Noheda, B. A Rhombohedral Ferroelectric Phase in Epitaxially-Strained $Hf_{0.5}Zr_{0.5}O_2$ Thin Films. *Nature Mater.* **2018**, 17, 1095-1100.

(15) Lyu, J.; Fina, I.; Solanas, R.; Fontcuberta, J.; Sánchez, F. Robust Ferroelectricity in Epitaxial $Hf_{1/2}Zr_{1/2}O_2$ Thin Films. *Appl. Phys. Lett.* **2018**, 113, 082902.

(16) Lee, K.; Lee, T. Y.; Yang, S. M.; Lee, D. H.; Park, J.; Chae, S. C. Ferroelectricity in Epitaxial Y-doped $HfO_2$ Thin Film Integrated on Si Substrate. *Appl. Phys. Lett.* **2018**, 112, 202901.

(17) Müller, J.; Polakowski, P.; Mueller S.; Mikolajick, T. Ferroelectric Hafnium Oxide Based Materials and Devices: Assessment of Current Status and Future Prospects. *ECS J. Solid State Sci. Technol.* **2014**, 64, 159-168.

(18) Ambriz-Vargas, F.; Kolhatkar, G.; Thomas, R.; Nouar, R.; Sarkissian, A.; Gomez-Yáñez, C.; Gauthier, M. A.; Ruediger, A. Tunneling Electroresistance Effect in a $Pt/Hf_{0.5}Zr_{0.5}O_2/Pt$ Structure. *Appl. Phys. Lett.* **2017**, 110, 093106.

(19) Ambriz-Vargas, F.; Kolhatkar, G.; Broyer, M.; Hadj-Youssef, A.; Nouar, R.; Sarkissian, A.; Thomas, R.; Gomez-Yáñez, C.; Gauthier, M. A.; Ruediger, A. A Complementary Metal Oxide Semiconductor Process-Compatible Ferroelectric Tunnel Junction. *ACS Appl. Mater. Interfaces* **2017**, 9, 13262-13268.

(20) Tian, X.; Shibayama, S.; Nishimura, T.; Yajima, T.; Migita, S.; Toriumi, A. Evolution of Ferroelectric $HfO_2$ in Ultrathin Region Down to 3 nm. *Appl. Phys. Lett.* **2018**, 112, 102902.

(21) Meyer R.; Waser, R. Dynamic Leakage Current Compensation in Ferroelectric Thin-Film Capacitor Structures. *Appl. Phys. Lett.* **2005**, 86, 142907.

(22) Fina, I.; Fàbrega, L.; Langenberg, E.; Martí, X.; Sánchez, F.; Varela, M.; Fontcuberta, J. Non-Ferroelectric Contributions to the Hysteresis Cycles in Manganite Thin Films: a Comparative Study of Measurement Techniques. *J. Appl. Phys.* **2011**, 109, 074105.

(23) Park, M. H.; Kim, H. J.; Kim, Y. J.; Lee, W.; Moon, T.; Hwang, C. S. Evolution of Phases and Ferroelectric Properties of Thin $Hf_{0.5}Zr_{0.5}O_2$ Films According to the Thickness and Annealing Temperature. *Appl. Phys. Lett.* **2013**, 102, 242905.

(24) Park, M. H.; Kim, H. J.; Kim, Y. J.; Moon, T.; Hwang, C. S. The Effects of Crystallographic Orientation and Strain of Thin $Hf_{0.5}Zr_{0.5}O_2$ Film on its Ferroelectricity. *Appl. Phys. Lett.* **2014**, 104, 072901.

(25) Park, M. H.; Kim, H. J.; Kim, Y. J.; Lee, Y. H.; Moon, T.; Kim, K. D.; Hyun, S. D.; Fengler, F.; Schroeder, U.; Hwang, C. S. Effect of Zr Content on the Wake-Up Effect in $Hf_{1-x}Zr_xO_2$ Films. *ACS Appl. Mater. Interf.* **2016**, 8, 15466-15475.

(26) Park, M. H.; Lee, Y. H.; Mikolajick, T.; Schroeder, U.; Hwang, C. S. Review and Perspective on Ferroelectric $HfO_2$-based Thin Films for Memory Applications. *MRS Commun.* **2018**, 8, 795-808.

(27) Clima, S.; Wouters, D. J.; Adelmann, C.; Schenk, T.; Schroeder, U.; Jurczak, M.; Pourtois. G. Identification of the Ferroelectric Switching Process and Dopant-dependent Switching Properties in Orthorhombic $HfO_2$: A First Principles Insight. *Appl. Phys. Lett.* **2014**, 104, 092906.

(28) Scott, J. F.; Kammerdinger, L.; Parris, M.; Traynor, S.; Ottenbacher, V.; Shawabkeh, V.; Oliver, W. F. Switching Kinetics of Lead Zirconate Titanate Submicron Thin film Memories. *J. Appl. Phys.* **1988**, 64, 787-792.

(29) Yan, H.; Inam, F.; Viola, G.; Ning, H.; Zhang, H.; Jiang, Q.; Zeng, T.; Gao, Z.; Reece, M. J. The Contribution of Electrical Conductivity, Dielectric Permittivity and Domain Switching in Ferroelectric Hysteresis Loops. *J. Adv. Dielectrics* **2011**, 1, 107-118.

(30) Pešić, M.; Fengler, F. P. G.; Larcher, L.; Padovani, A.; Schenk, T.; Grimley, E. D.; Sang, X.; LeBeau, J. M.; Slesazeck, S.; Schroeder, U.; Mikolajick, T. Physical Mechanisms behind the Field-Cycling Behavior of $HfO_2$-Based Ferroelectric Capacitors. *Adv. Funct. Mater.* **2016**, 26, 4601-4612.

# Supporting Information

# Epitaxial Integration on Si(001) of Ferroelectric $Hf_{0.5}Zr_{0.5}O_2$ Capacitors with High Retention and Endurance

*Jike Lyu, Ignasi Fina, Josep Fontcuberta, Florencio Sánchez\**

Institut de Ciència de Materials de Barcelona (ICMAB-CSIC), Campus UAB, Bellaterra 08193, Barcelona, Spain

**XRD 2θ-χ frames**

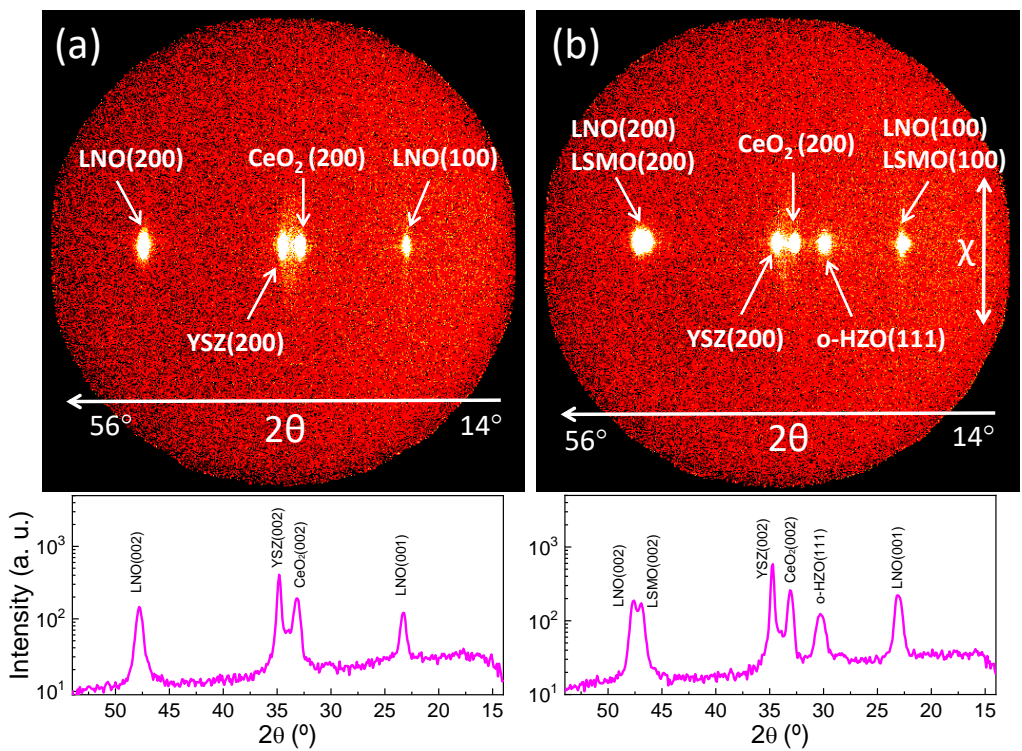

**Figure S1**: XRD 2θ-χ frames, recorded with a 2D detector, corresponding to (a) HZO/LNO/CeO$_2$/YSZ/Si(001) and (b) HZO/LSMO/LNO/CeO$_2$/YSZ/Si(001). The evident HZO(111) reflection in (b) cannot be appreciated in (a). The corresponding θ-2θ patterns (bottom panels) were obtained by integration along χ from -5 to +5°.

**XRD θ-2θ scan (long acquisition time) and glazing incidence XRD scan**

Figure S2 shows the XRD θ-2θ scan of the HZO/LSMO/LNO/CeO$_2$/YSZ/Si(001) sample measured with longer acquisition time per step than Figure 1 b. The pattern around HZO(111) and CeO$_2$(002) reflections has been simulated (red and blue curves, respectively) according the dependence[1]:

$$I(Q) = \left(\frac{\sin\left(\frac{QNc}{2}\right)}{\sin\left(\frac{Qc}{2}\right)}\right)^2$$

where $Q = 4\pi\sin(\theta)/\lambda$ is the reciprocal space vector, N the number of unit cells along the out-of-plane direction and c the corresponding lattice parameter. The simulation around HZO(111), red curve, has been fitted supposing a thickness of 97.3 Å (N = 33 and c = 2.948 Å), being the corresponding Bragg reflection located at 2θ = 30.319°. For the blue curve, corresponding to CeO$_2$(002), thickness was 168 Å (N = 31 and c = 5.41 Å), being the corresponding Bragg reflection located at 2θ = 33.117°.

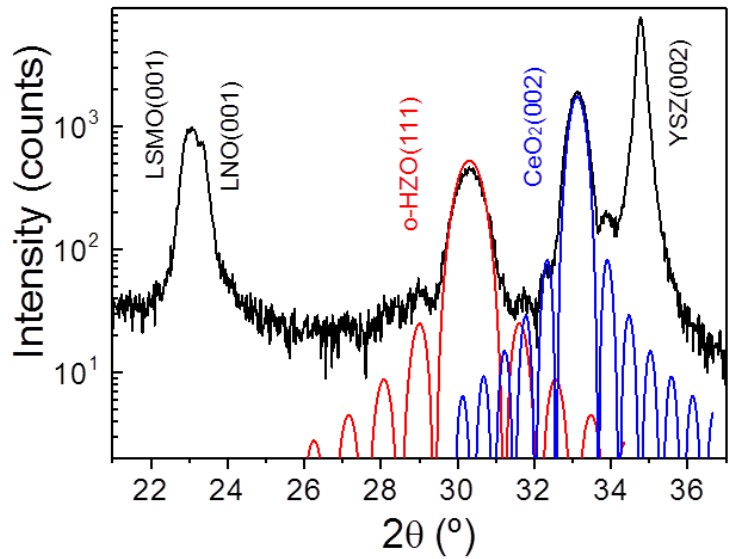

**Figure S2**: XRD θ-2θ scan of the HZO/LSMO/LNO/CeO$_2$/YSZ/Si(001) sample measured with long acquisition time.

**Grazing incidence XRD scan of the HZO/LNO/CeO$_2$/YSZ/Si(001) sample.**

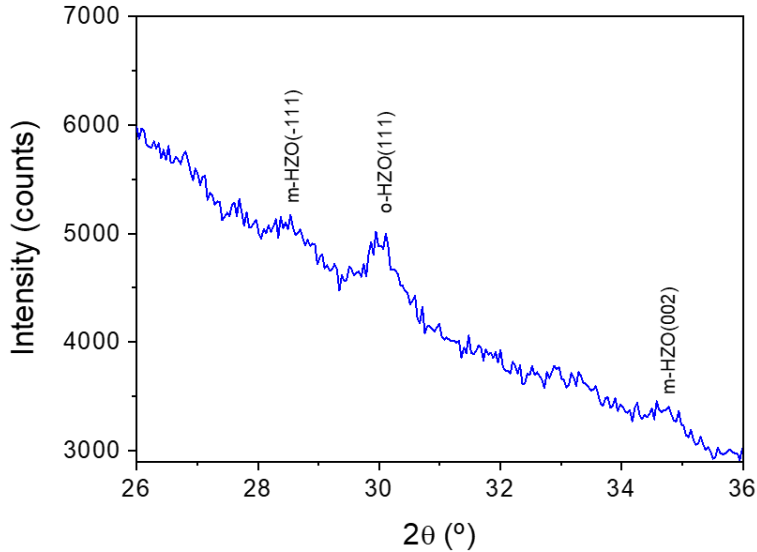

**Figure S3**: Grazing incidence XRD corresponding to the HZO/LNO/CeO$_2$/YSZ/Si(001) sample.

**Hysteresis loops recorded for increasing voltage**

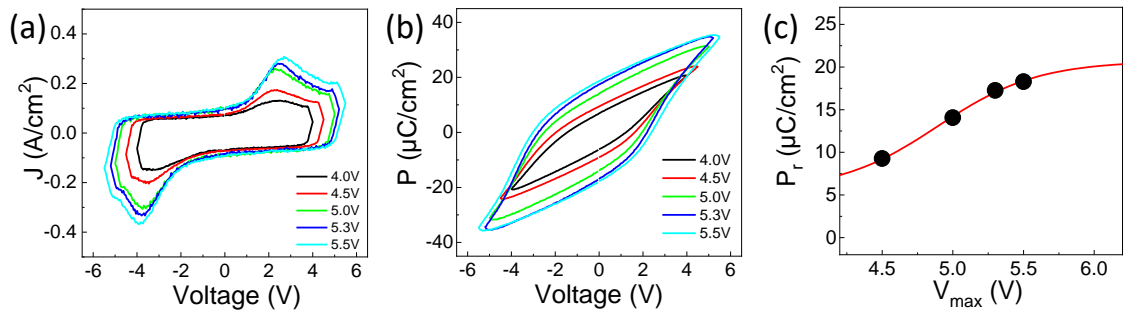

**Figure S4**: (a) Current versus voltage hysteresis loops recorded at 1 kHz for increasing voltage in the HZO/LSMO/LNO/CeO$_2$/YSZ/Si(001) sample. (b) Ferroelectric polarization versus voltage hysteresis loop recorded at 1 kHz measured for increasing voltage. It can be inferred that the loops measured at 5.3 and 5.5 V are very similar revealing that at 5.5 V the polarization is almost saturation. (c) P$_r$ versus maximum applied voltage. It can be observed that the P$_r$ value is gradually increasing until saturation. Measurements performed at larger applied voltage result in dielectric breakdown.

**Compensation of the leakage contributions in polarization loops**

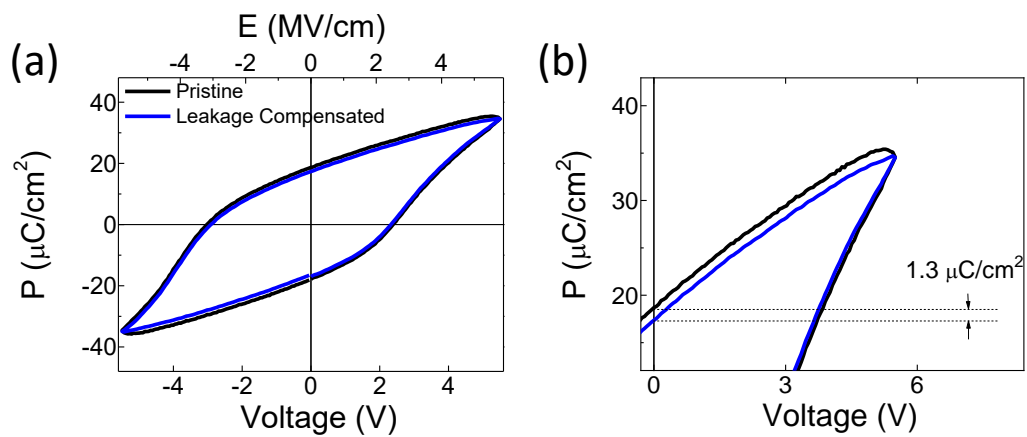

**Figure S5**: Compensation of the leakage contributions. (a) Pristine loop and compensated loop, showing very similar polarization values. (b) Zoom of data shown in (a), where it can be inferred that the leakage contribution produces an overestimation of P of near 1.3 $\mu C/cm^2$. The leakage current contribution has been estimated from exponential fitting at high voltage.

**Polarization hysteresis loop measured in the HZO/LNO/CeO$_2$/YSZ/Si(001)**

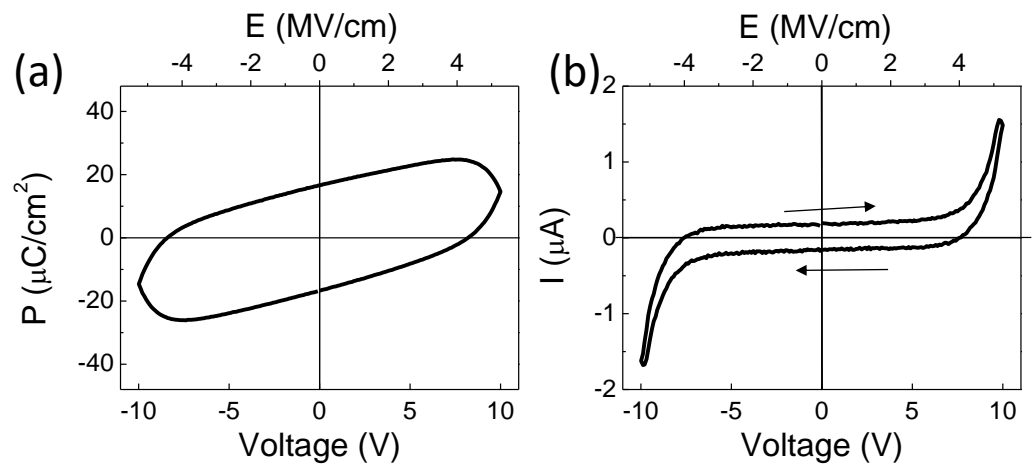

**Figure S6**: (a) Ferroelectric polarization versus voltage hysteresis loop recorded at 1 kHz and measured in the HZO/LNO/CeO$_2$/YSZ/Si(001) sample up to 10 V in top-top configuration. Cigar-shape like loop is observed without signature of ferroelectric switching. (b) Current versus voltage hysteresis loop recorded at 1 kHz. Non-zero current at low voltage range corresponds to displacive current due to the dielectric nature of the film. Current increase near the maximum applied voltage is due to leakage.

**Ferroelectric retention**

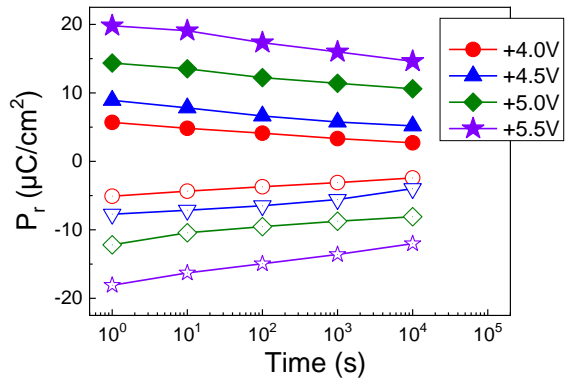

**Figure S7**: Ferroelectric retention in the HZO/LSMO/LNO/CeO$_2$/YSZ/Si(001) sample after positive (solid symbols) and negative (empty symbol) poling voltage pulses of indicated amplitude. It can be observed that irrespectively of the sign of the poling voltage, retention is very similar. This fact illustrates that the imprint electric field revealed by Figure 2a, which points towards LSMO electrode and which might favor positive polarization (towards LSMO), does not have important impact on ferroelectric retention of the device.

REFERENCES

(1) Pesquera, D.; Martí, X.; Holy, V.; Bachelet, R.; Herranz, G.; Fontcuberta, J. X-ray interference effects on the determination of structural data in ultrathin La$_{2/3}$Sr$_{1/3}$MnO$_3$ epitaxial thin films. *Appl. Phys. Lett.* **2011**, 99, 221901.